\documentclass[fleqn,10pt]{wlscirep}
\usepackage[utf8]{inputenc}
\usepackage[T1]{fontenc}
\title{Human-centered evaluation of data visualization understanding in vision-language models}

\author[1,2,*]{Arnav Verma}
\author[1]{Kushin Mukherjee}
\author[1]{Christopher Potts}
\author[3]{Elisa Kreiss}
\author[1,+]{Judith~E.~Fan}
\affil[1]{Stanford University, USA}
\affil[2]{Massachusetts Institute of Technology, USA}
\affil[3]{University of California, Los Angeles, USA}

\affil[*]{arnavv@mit.edu}
\affil[+]{jefan@stanford.edu}

\usepackage{mathptmx}                  
\usepackage{times,latexsym}
\usepackage{url}
\usepackage[T1]{fontenc}
\usepackage{graphicx}
\usepackage{titlesec}
\titlespacing*{\paragraph}{\parindent}{0.25ex}{1ex}
\usepackage[svgnames]{xcolor}
\usepackage{float}
\usepackage{booktabs}
\usepackage{tabularx}
\usepackage{array}
\definecolor{FlanXLColor}{HTML}{5ba3cf}
\newcommand{\flanXL}{\textcolor{FlanXLColor}{\textbf{Blip2-FlanT5-4B}}}

\definecolor{FlanXXLColor}{HTML}{4c78a8}
\newcommand{\flanXXL}{\textcolor{FlanXXLColor}{\textbf{Blip2-FlanT5-11B}}}

\definecolor{llavaSevenColor}{HTML}{f9b574}
\newcommand{\llavaSeven}{\textcolor{llavaSevenColor}{\textbf{LLaVA1.5-Vicuna-7B}}}

\definecolor{llavaThirteenColor}{HTML}{f58518}
\newcommand{\llavaThirteen}{\textcolor{llavaThirteenColor}{\textbf{LLaVA1.5-Vicuna-13B}}}

\definecolor{llavaThirtyFourColor}{HTML}{BF9000}
\newcommand{\llavaThirtyFour}{\textcolor{llavaThirtyFourColor}{\textbf{LLaVA1.6-Yi-34B}}}

\definecolor{pixColor}{HTML}{b9a7d0}
\newcommand{\pix}{\textcolor{pixColor}{\textbf{Pix2Struct-0.3B}}}

\definecolor{matchaChartqaColor}{HTML}{8b6db2}
\newcommand{\matcha}{\textcolor{matchaChartqaColor}{\textbf{MatCha-0.3B}}}

\definecolor{gptvColor}{HTML}{b85536}
\newcommand{\gptv}{\textcolor{gptvColor}{\textbf{GPT-4V}}}

\definecolor{humansColor}{HTML}{2e693b}
\newcommand{\humans}{\textcolor{humansColor}{\textbf{Humans}}}
\usepackage{colortbl}
\usepackage{soul}

\newcommand{\ci}[3]{#1; 95\%~CI~=~[#2,~#3]}

\newcommand{\revtwo}[1]{#1}

\keywords{Chart understanding, Graph comprehension, Visualization literacy, Artificial intelligence, Benchmarking}

\begin{abstract}
{
Data visualization understanding, which requires integrating visual, numerical, and linguistic inputs, poses fundamental challenges to multimodal reasoning. 
Recently developed vision-language models (VLMs) are promising candidates for modeling the computations that underlie human data visualization understanding. 
However, whether they are useful as scientific models depends not only on whether they produce correct answers, but also whether they approximate the patterns of responses that people produce on the same tasks. 
This distinction has been ignored in recent evaluations of VLMs, which typically only report overall model performance (accuracy) without direct comparison to human behavioral \textit{patterns}. Here we introduce a human-centered evaluation protocol that compares VLM and human responses on the same set of visualization understanding items. We apply this protocol to eight VLMs across six assessments: five established tests of human visualization literacy, plus one designed specifically for evaluating VLMs. Models performed worse than humans on average, and this gap persisted even under lenient criteria. Moreover, while relative performance across items was somewhat correlated between models and humans, all models produced error patterns that were reliably distinct from those of humans. We envision our evaluation protocol helping to identify artificial systems that might serve as promising scientific cognitive models of how humans reason about data visualizations.

}

\end{abstract}
\begin{document}

\flushbottom
\maketitle

\thispagestyle{empty}


\section*{Introduction}
\begin{figure*}[ht!]
    \centering
    \includegraphics[width=\linewidth]{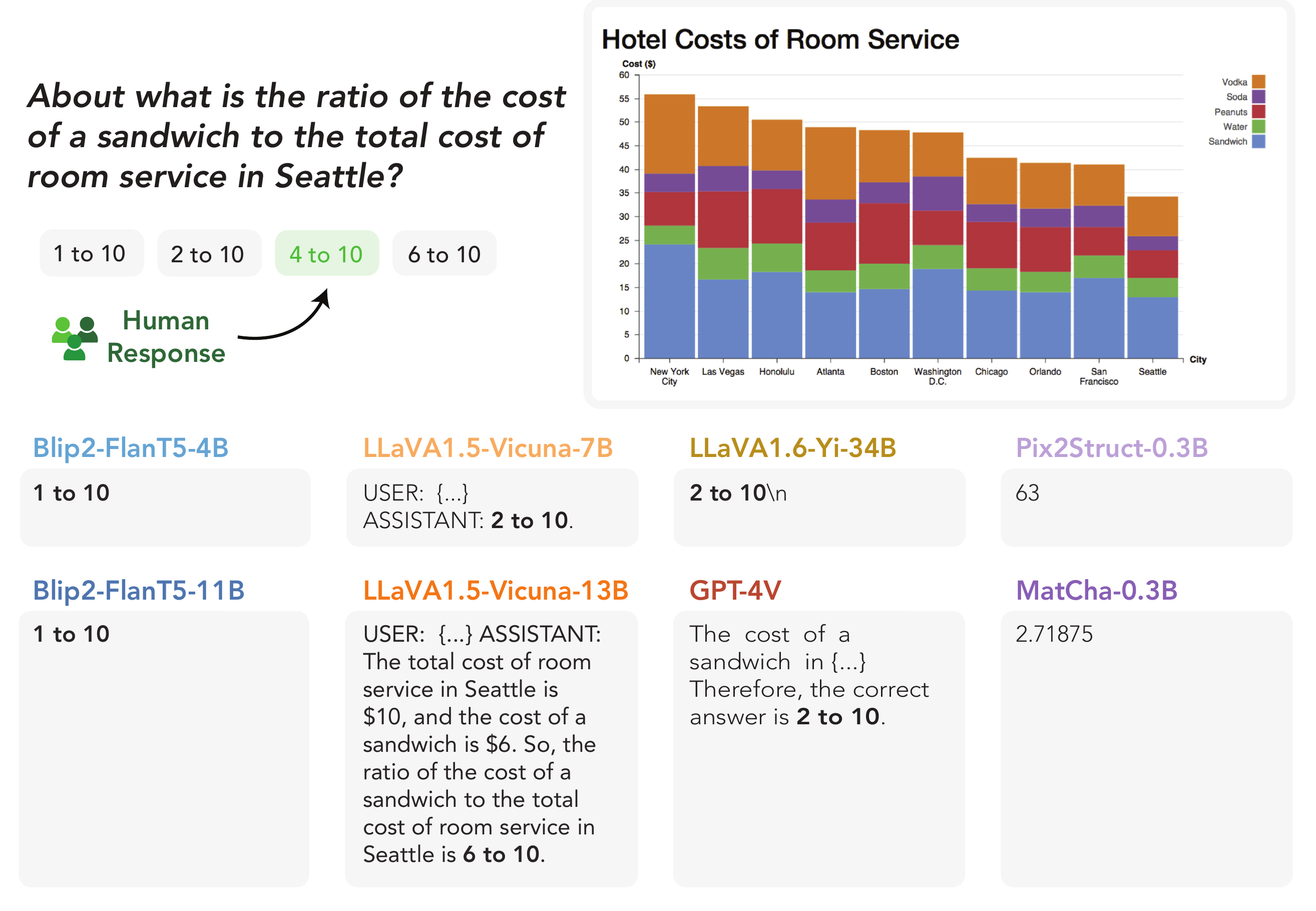}
    \caption{
    Sample response from all evaluated models for a multiple-choice item.  Responses after processing are shown in bold and are used for comparison against human and model responses. Responses without bold characters indicate invalid responses.} 
    \label{fig:response-examples}
\end{figure*}

Humans can engage with a wide range of visual input modalities, ranging from natural scenes and drawings to diagrams and data visualizations \cite{tversky2011visualizing,franconeri2021science, fan2023drawing}. 
Data visualizations --- also commonly known as \textit{graphs}, \textit{charts}, and/or \textit{plots} --- are indispensable tools for supporting exploratory analysis and statistical reasoning \cite{tukey1977exploratory,borner2019data,cumming2005inference}. 
They do so by leveraging a combination of visual features (e.g., color, shape, size, position) and text-based annotations (e.g., axis labels, legends) to efficiently convey patterns in quantitative data \cite{bertin1981graphics, tufte1983visual,wilkinson2012grammar, munzner2014visualization}.
As such, interpreting any data visualization relies upon the ability to correctly combine visual, linguistic, and quantitative information to answer some question at hand (Figure \ref{fig:response-examples}).
Moreover, the acquisition of data visualization literacy --- a robust ability to parse data visualizations and derive insights from them \cite{fry1981graphical, curcio1987comprehension, friel2001making, shah2002review,  boy2014principled,borner2019data,firat2022interactive} --- has been a longstanding priority in STEM education \cite{national2014developing, carver2016guidelines}.

Nevertheless, there are fundamental gaps in current knowledge of what cognitive operations underlie data visualization understanding. 
In part, these gaps reflect inherent challenges in operationalizing such a complex cognitive construct.
The complexity stems from the fact that data visualizations can be made to facilitate understanding of many different quantitative phenomena (e.g., a person might sometimes want only to search for a single value and other times to derive broader insights about complex trends), and the goals of the visualization designer can lead to the same dataset being visualized in many different ways \cite{brehmer2013multi, friel2001making, amar2005low, quadri_survey_2022}.
The primary strategy for enhancing understanding of data using visualization is to encode the underlying data using different visual channels (size, shape, color, etc.) in order to produce different types of data visualizations (bar plots, line plots, scatter plots, etc.) \cite{boy2014principled,lee2016vlat,kim2018assessing, borner2019data, lundgard2021accessible, quadri_survey_2022}.
The ability to perform visualization understanding tasks is thought to rely on the coordination of several mental processes \cite{hegarty2005multimedia}, including: rapid perceptual computations \cite{cleveland1984graphical, schloss2025perceptual} with respect to a known graph schema \cite{pinker1990theory}; explicit numerical operations \cite{gillan1994componential} constrained by finite working memory resources \cite{padilla2018decision}; and interpretive processes that lead to more general insights \cite{carpenter1998model}, which may be influenced by prior content knowledge \cite{shah2011bar, xiong2019curse}.

Classical accounts of these processes are limited in that they either are not specified in computationally explicit terms or are derived based on a limited variety of data visualizations, thus limiting their generalizability \cite{pinker1990theory, shah2005comprehension, simkin1987information, fox2023theories}. 
To more precisely describe the operations that enable visualization understanding, as well as developmental changes accompanying the acquisition of data visualization literacy, there is a need for computational models that can contend with the diversity of real-world visualizations and are adaptable to common visualization understanding tasks. What is needed are \emph{cognitive models} --- computational artifacts that take the same inputs as a person, produce outputs that match human behavior, and that can be used to both predict and explain it\cite{frank2025cognitive}.
Recently developed ``multimodal'' AI systems are promising candidate models because they can operate over a combination of visual and textual inputs to perform a wide variety of tasks that require integrating information from both these channels \cite{openai2023gpt4, alayrac2022flamingo, laurenccon2024matters}.
The complexity of tasks that these systems have been reported to perform well has begun to approach that of tasks that humans routinely face in real-world settings, including at school and in the workplace \cite{chung2024scaling, zhang2022opt,openai2023gpt4, liu2024improved, bommasani2021opportunities,katz2023gpt, yue2024mmmu}.
This progress has made it plausible to consider vision-language models (VLMs) as providing a foundation for developing scientific models of \emph{human reasoning} over multiple information modalities.

\begin{figure*}[ht!]
  \centering\includegraphics[width=\linewidth, alt={A view of a city with buildings peeking out of the clouds.}]{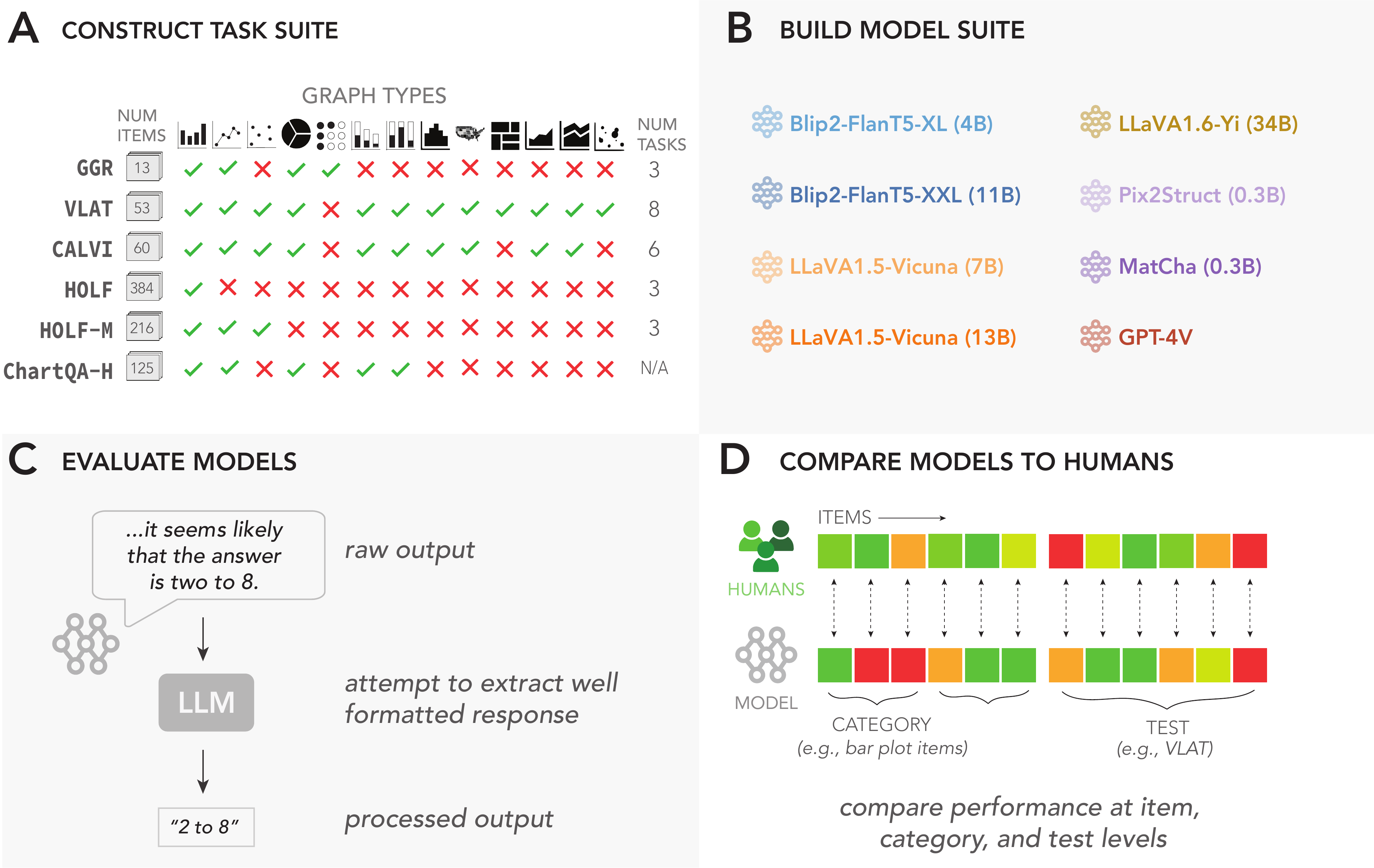}
  \caption{Our evaluation protocol uses six assessments to evaluate a suite of eight vision-language models. Model performance is compared to human performance across different item, category, and test levels.
  }
  \label{fig:teaser}
\end{figure*}

However, for such AI systems to provide a useful basis for developing cognitive models of human visualization understanding, it is first critical to evaluate to what degree they generate patterns of behavior on data visualization understanding tasks that approximate those generated by humans.
While strong accuracy has been reported for some of these systems on data visualization understanding tasks, these reports rely upon different measures from those typically used to assess the same abilities in humans and generally do not directly compare models' behavior to that of humans\cite{masry2022chartqa,lu2023mathvista,openai2023gpt4, yue2024mmmu, methani2020plotqa, wang2024charxiv, wu2024chartinsights, tang2025chartmuseum, mukherjee2025encqa}. 
As such, it remains unclear to what degree any current systems approach human-level abilities or engage in human-like reasoning about data visualizations, thereby limiting any insights that can be drawn about the operations involved in human visualization understanding from such models.
Our paper addresses this gap in three ways (Figure~\ref{fig:teaser}):

\textit{First}, we present \texttt{CHART-6} (\textit{\textbf{C}omparative \textbf{H}uman-\textbf{A}I graphical \textbf{R}easoning \textbf{T}ests}), a human-centered suite of data visualization understanding assessments from the psychology and visualization literature (Figure~\ref{fig:teaser}A: Construct Task Suite). This suite contains three assessments commonly used to evaluate data visualization literacy in humans (GGR~\cite{galesic2011graph}, VLAT~\cite{lee2016vlat}, and CALVI~\cite{ge2023calvi}); two procedurally generated assessments (HOLF~\cite{huey2023communicative} and HOLF-Multi~\cite{huey2024adaptive}), which are employed in psychological experiments on  communication through data visualizations; and a subset of a popular assessment for assessing data visualization understanding in VLMs (ChartQA-Human~\cite{masry2022chartqa}).

\textit{Second}, we develop an evaluation protocol to rigorously assess the performance of vision-language models on question-answering tasks germane to visualization understanding, designed to enable direct comparison to human behavior (Figure~\ref{fig:teaser}C: Evaluate Models). 

\textit{Third}, we use this protocol to evaluate the performance of several vision-language models (Figure~\ref{fig:teaser}B: Build Model Suite) against human behavior on \texttt{CHART-6}, with respect to both how well these models perform and how well they emulate \textit{human-like} behavior on these tests (Figure~\ref{fig:teaser}D: Compare Models to Humans).

 Our evaluations revealed that many of these models often failed to produce valid responses when administered these tests.
Even when focusing on items for which models did produce valid responses, we found that they achieved reliably lower performance than did the adult human participants represented in this work.
Direct comparison of human and model response patterns revealed that while there were some categories of items (e.g., visualizations using bar graphs, questions requiring retrieving specific values from visualizations) where human and model performance was comparable.
We further found that humans generally outperformed models in this evaluation, and that the items which humans found difficult were not necessarily those on which these models also displayed worse performance.

Nevertheless, we found that no model produced patterns of responses that approached the human noise ceiling.
 While we take these findings as suggestive that further innovation is needed to develop models that can form the basis of cognitive models of human visualization understanding,  we wish to emphasize that the results we report in this study represent a point-in-time comparison, and that we expect that evaluations will need to be conducted on an ongoing basis in order to accurately track progress towards this end. 
We envision our work as contributing to the development of protocols for comparing human and model patterns of behavior on multimodal reasoning tasks, by grounding these evaluations in well-validated human-centered assessments.

\section*{Results}

 We collated and standardized the set of test items (i.e., unique visualization-question pairs) drawn from the six assessments that constitute \texttt{CHART-6}. The final dataset consisted of 851 unique items spanning different visualization and question types (see Methods).
We leveraged existing human behavioral data for five of the six tests, and collected additional human behavioral responses for a representative subset of the ChartQA benchmark.

Next, we generated responses from eight state-of-the-art vision–language models for each test item using a standardized protocol.
Our evaluation protocol proceeded through the following stages: \textit{first}, we assessed whether models produced valid responses. 
\textit{Second}, we evaluated their accuracy against the ground-truth responses provided by the original test designers. 
\textit{Lastly}, we compared model and human responses at two levels of granularity: first, grouped by the type of visualization (e.g., bar graph, line graph, pie chart) and question (e.g., value identification, arithmetic computation), and second, at the level of individual test items.

\subsection*{How often do models produce \textit{valid} responses?} 
\begin{figure*}[ht!]
    \centering
    \includegraphics[width=0.99\linewidth]{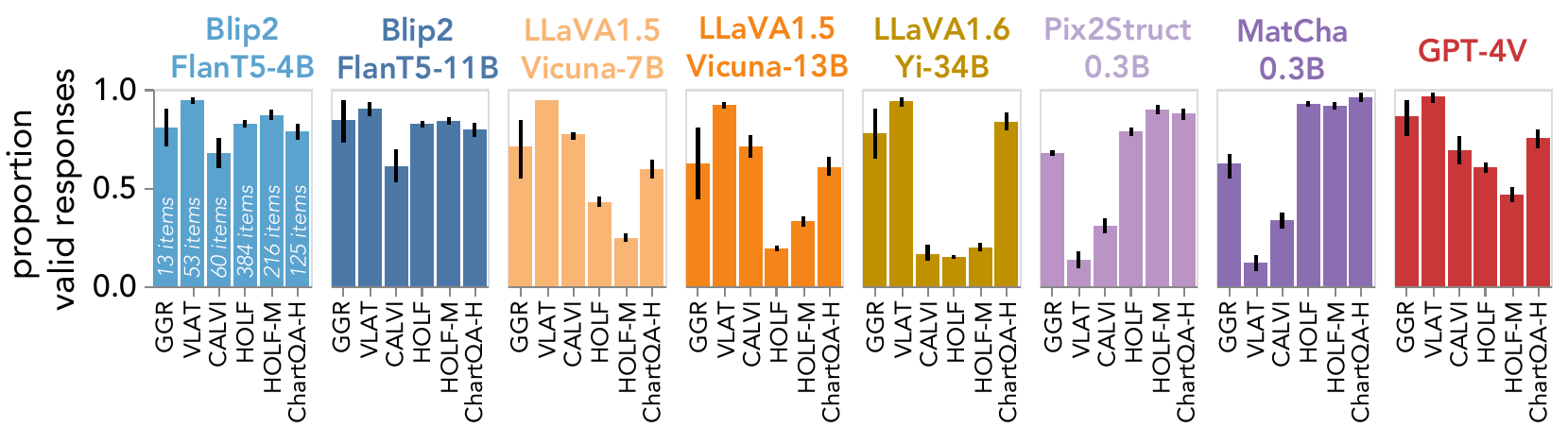}
    \caption{Proportion of valid responses produced by each model on each assessment.}
    \label{fig:valid_responses}
\end{figure*}

 Before comparing human and model response patterns, we first assessed to what extent this suite of VLMs could achieve reasonable levels of performance on the data visualization understanding tasks in \texttt{CHART-6}. 
We evaluated models using two standards. First, we evaluated their ability to produce correctly formatted, `valid' responses to the questions being posed. Second, we evaluated their ability to produce answers that are correct, as defined by the original test designers.

Thus, for each of the eight vision-language models in our suite, we  determined the proportion of model responses that were at least correctly formatted, and thus amenable to further analysis.
For multiple-choice questions in tests GGR, VLAT, and CALVI, a response was considered to be valid if the processed response was an exact match to one of the multiple-choice options.
For numerical-response questions in tests GGR, HOLF, HOLF-Multi, and ChartQA-Human, a response was considered to be valid if it contained a single floating-point value.

Using these criteria, we computed the proportion of valid responses generated for each test item by every model (Figure \ref{fig:valid_responses}).
We found that no model consistently provided valid responses.
When pooling all items across tests, we found that \llavaThirtyFour{} produced the lowest proportion of valid responses (average: \ci{0.32}{0.30}{0.34}; 2735/8510 responses were valid) and \matcha{} produced the highest proportion of valid responses (average: \ci{0.83}{0.82}{0.84}; 7082/8510 responses were valid). 
However, the BLIP-2 variants stood out for most consistently producing a high proportion of valid responses across all tests (\flanXL{}: \ci{0.82}{0.75}{0.88}); \flanXXL{}: \ci{0.80}{0.72}{0.86}).

These results suggest that reliably extracting task-relevant output from VLMs remains challenging.
This limitation has implications for the way that sound comparisons between model and human performance can be made, depending on whether invalid responses are considered to be incorrect responses generated under fair evaluation settings, and thus reflect limitations of the model, or are considered to be the product of limitations in our evaluation protocol.
To ensure that our conclusions are not dependent on this choice, we conducted subsequent analyses under both ways of interpreting invalid responses from models. 

\subsection*{How often do models produce \textit{accurate} answers?} 

\begin{figure*}[ht!]
    \centering
\includegraphics[width=\linewidth]{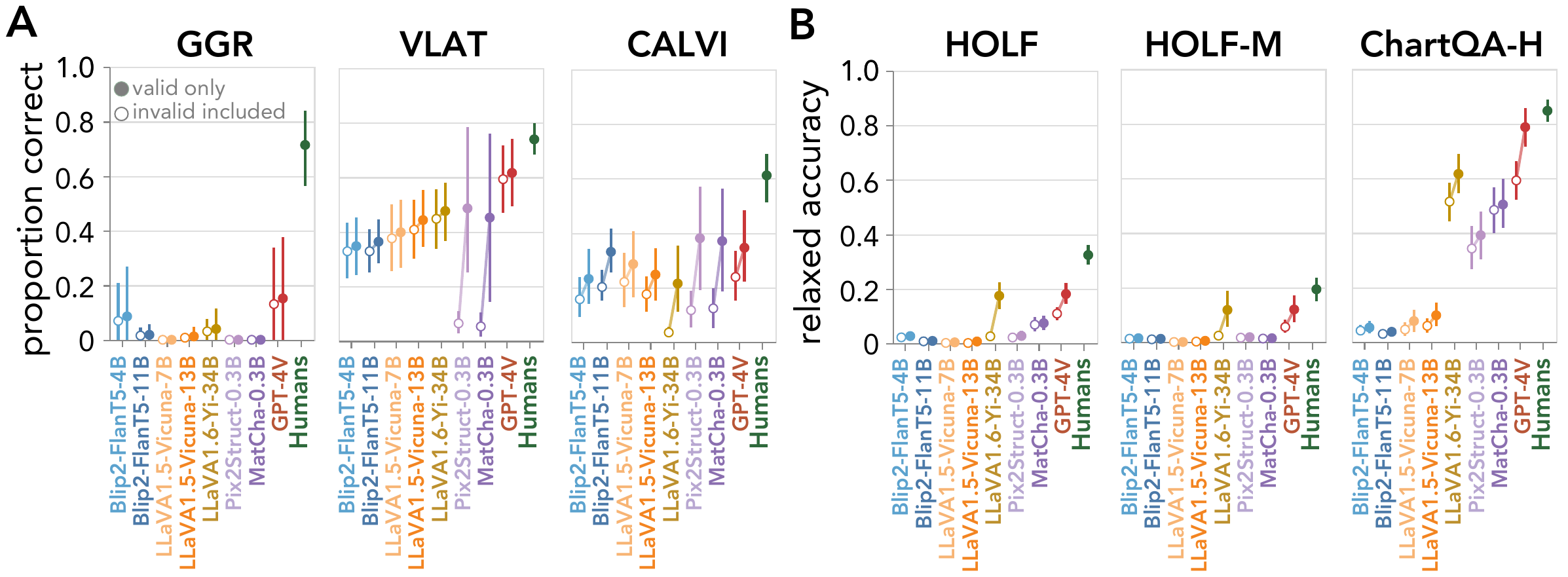}
\caption{Human and model performance on (A) the mean proportion correct in multiple-choice assessments (GGR, VLAT, and CALVI) and (B) the mean relaxed accuracy in numerical-response assessments (HOLF, HOLF-Multi, and ChartQA). Relaxed accuracy is calculated by the proportion of responses that fall within 5\% of the correct answer. Empty circles represent estimates of model performance based on all responses, with any invalid responses marked as incorrect. Filled circles represent estimates of model performance based only on valid responses, and therefore reflect an upper bound on model performance. All error bars represent bootstrapped 95\% confidence intervals.}
\label{fig:agent_accuracy}
\end{figure*}

Next, we compared the accuracy of the responses achieved by models to that of the average human participant who completed the same six tests in \texttt{CHART-6} (Figure \ref{fig:agent_accuracy}). 
We established an upper and lower bound on estimates of model performance by computing accuracy when considering only valid responses (upper bound) and when considering all responses, including invalid ones, where invalid responses were marked as incorrect (lower bound).
For GGR, VLAT, and CALVI (the `multiple-choice' tests), we computed the proportion of correct responses produced by humans and models. 
For the 9 items requiring numerical responses in GGR, responses were only deemed correct if they \textit{exactly} matched the ground-truth answer provided by the original test designers, to ensure fair comparisons between vision-language models and human responses to items on this test.
For HOLF, HOLF-multi, and ChartQA (the `numerical-response' tests), following prior work \cite{masry2022chartqa,methani2020plotqa}, we computed a \textit{relaxed accuracy} metric, which considers a response to be correct if it falls within 5\% of the correct answer. 
The same standard was applied to both human and vision-language model responses.

We found that models reliably differed in performance from one another ($\chi^{2}(7)$~=~3,591, $p$~<~0.001). 
We further found that considering only valid responses inflated estimates of model performance on the numerical-response tests to some degree ($\Delta$ proportion correct: \ci{0.041}{0.024}{0.057}), with a more modest impact on estimates of model performance on multiple-choice tests ($\Delta$ proportion correct: \ci{0.12}{0.066}{0.18}).
These results suggest the value of jointly considering both stricter and more lenient ways of assessing model performance to more clearly establish the range of expected performance levels for any given model.

When examining only the valid responses generated by models (Figure \ref{fig:agent_accuracy}), we found that \gptv{} was the best performing model on five out of the six tests. 
However, it performed reliably worse than human participants on GGR ($\Delta$ mean proportion correct (model $-$ human): \ci{-0.56}{-0.78}{-0.30}), HOLF ($\Delta$ mean relaxed accuracy: \ci{-0.15}{-0.25}{-0.14}), and HOLF-Multi ($\Delta$ mean relaxed accuracy: \ci{-0.07}{-0.16}{-0.02}). 
It did approach human performance on VLAT ($\Delta$ mean proportion correct: \ci{-0.12}{-0.26}{0.01}) and ChartQA-Human ($\Delta$ mean relaxed accuracy: \ci{-0.060}{-0.14}{0.020}).
\pix{} performed best among models on CALVI approaching human performance ($\Delta$ mean proportion correct: \ci{-0.23}{-0.42}{-0.018}). 
Among those items for which \pix{} could generate a valid response at all, the gap between \pix{} and \humans{} was all but closed for the misleading items  ($\Delta$ mean proportion correct: \ci{-0.01}{-0.28}{0.22}), but not for the non-misleading items ($\Delta$ mean proportion correct: \ci{-0.42}{-0.90}{-0.22}).

We also compared model performance to that of humans using all model responses, with any invalid model responses marked as incorrect. 
Under these conditions, we found that \gptv{} performed best among all models on all six assessments, including CALVI.
Again we found that \gptv{} performed worse than human participants on several of the tests: GGR ($\Delta$ mean proportion correct (model $-$ human): \ci{-0.58}{-0.76}{-0.35}), CALVI ($\Delta$ mean proportion correct: \ci{-0.37}{-0.50}{-0.23}), HOLF ($\Delta$ mean relaxed accuracy: \ci{-0.37}{-0.50}{-0.23}), and HOLF-Multi ($\Delta$ mean relaxed accuracy: \ci{-0.14}{-0.19}{-0.09}).
It did approach human-level performance on VLAT ($\Delta$ mean proportion correct: \ci{-0.12}{-0.25}{0.01}). 
However, by contrast with what we found when examining only valid responses, \gptv{} did not achieve human-level performance on ChartQA-Human ($\Delta$ mean relaxed accuracy: \ci{-0.26}{-0.34}{-0.17}). 

In particular, when we considered only items for which models produced valid responses, thus potentially overestimating models' performance, the model–human performance gap narrowed substantially.
The effect was particularly pronounced for items from ChartQA, a dataset widely used to benchmark multimodal reasoning in frontier AI systems \cite{openai2023gpt4,team2024gemini,grattafiori2024llama}.
However, this gap widened when we considered all responses generated by models, including those on items for which it never produced a properly formatted response.
Taken together, our analyses show a reliable gap in performance between these models and human participants on several of the tests in our suite, including GGR, CALVI and HOLF, suggesting the value of using a diversity of independently designed measures for identifying opportunities for further model development.

\subsection*{How does model and human performance vary across different types of graphs and tasks?}

\begin{figure*}[ht!]
    \centering
    \includegraphics[width=\linewidth]{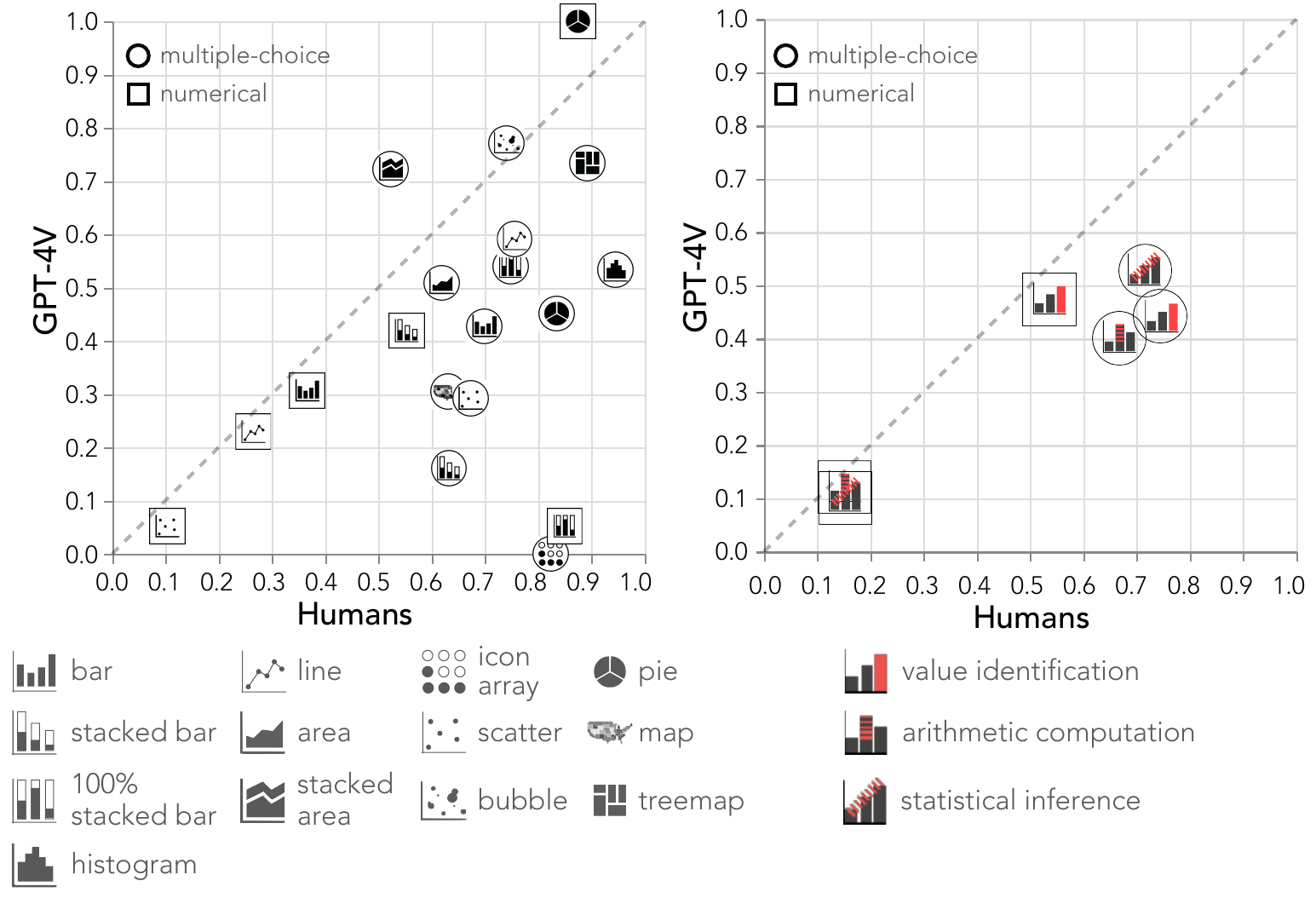}
    \caption{Mean proportion of correct responses between \gptv{} and \humans{} across different categories of \emph{graphs} (left) and \emph{tasks} (right).}
    \label{fig:feature_performance}
\end{figure*}

We next examined the degree to which the model-human performance gap varied across different categories of test items, regardless of which test they had come from.
Specifically, we examined variation in performance that could be attributable to the type of graph shown (e.g., bar plot vs. scatter plot) or the type of task being performed (e.g., value identification vs. arithmetic computation). 
Here we focused on the comparison between \gptv{} and \humans{}, because \gptv{} was the most consistently high-performing model on our suite of tests. 

We found that \humans{} consistently outperformed \gptv{} on most types of graphs (Figure~\ref{fig:feature_performance} left). 
The exceptions were stacked area and bubble charts that required multiple-choice responses and pie charts that required numerical responses. 
For these categories of items, \gptv{} scored higher than humans.
Qualitatively, while it seemed that the kinds of graphs on which \humans{} performed well were also the ones on which \gptv{} performed well, we did not find statistical support for this relationship (Pearson's $r (17) =0.40$, $p$~=~0.091). 
Indeed, we found that the size of the gap between \gptv{} and \humans{} varied across different types of graphs ($\chi^{2}(12)$~=~23.993, $p$~=~0.020).

We found that \humans{} outperformed \gptv{} across all three task types (i.e., value identification, arithmetic computation, statistical inference; Figure~\ref{fig:feature_performance} right). 
We further found the kinds of tasks on which \humans{} performed well were also often the ones on which \gptv{} also performed well (Pearson's $r(4)=0.94$, $p$~=~0.005), though the size of the gap did vary across tasks ($\chi^{2}(2)$~=~21.288, $p$~<~0.001).

We noticed that among the graph types where the human-model gap was largest were stacked bar plots and icon arrays (Figure~\ref{fig:feature_performance}, left). 
In addition, \gptv{} seemed to struggle more than \humans{} on items requiring multiple-choice responses, although the reasons this is the case remain unclear from these data alone. 
We believe these analyses provide a principled basis for identifying which data visualization understanding subtasks were especially difficult even for the most performant model in our evaluation, and thus where further model development might be most usefully directed.


\subsection*{How \textit{similar} were the error patterns generated by models and humans?} 
\begin{figure*}[ht]
    \centering
    \includegraphics[width=\linewidth]{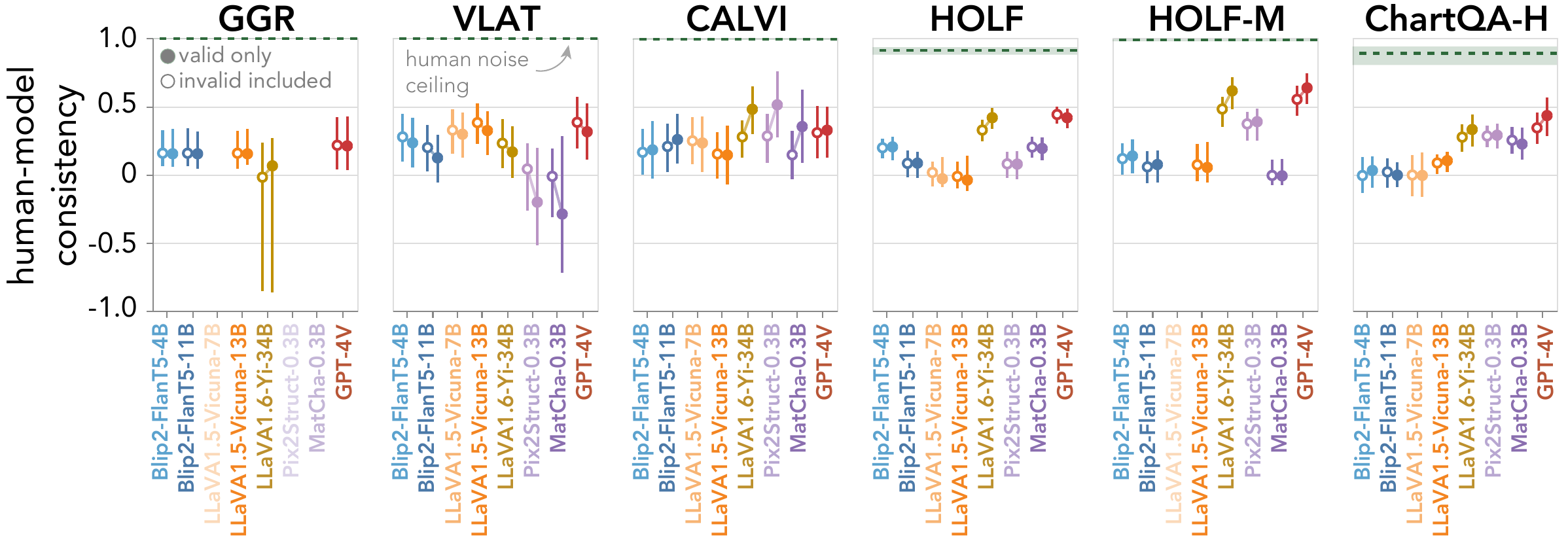}
    \caption{Pearson correlation between error patterns produced by each model and \humans{} on the same assessments. Empty circles represent estimates of model-human correlation based on all responses, with any invalid responses marked as incorrect. Filled circles represent estimates of model-human correlation based only on valid responses. All error bars represent bootstrapped 95\% confidence intervals. Human noise ceiling constructed by estimating the sampling distribution of the Spearman-Brown-corrected correlations between split-halves of the human data.}
    \label{fig:agent_consistency_err}
\end{figure*}

To more thoroughly investigate any covariation between human and model performance, we analyzed the full set of error patterns produced by humans and all models across the six assessments (Figure~\ref{fig:agent_consistency_err}).
Comparing error patterns across items is valuable because they could reveal aspects of how model and human behavior relate to each other that might not be apparent based on analyses of average performance on entire tests or pre-defined categories of items.
For each model, we constructed two versions of an error-pattern vector with length equal to the total number of items, where each element represented the proportion of correct responses it generated for a single item. 
One version of this error-pattern vector was derived from those items for which that model had generated at least one valid response; the second version was defined for all items, including ones where the model generated only invalid responses, which were marked as incorrect.
We then constructed an analogous error-pattern vector for humans, where each element represented the proportion of correct responses across all participants who had been given that item. 
Next, we computed how correlated the error-pattern vectors were between humans and all models. 
We computed a human ``noise ceiling'' reflecting how well any model could be expected to approximate human error patterns, given the variability in our estimates of human performance. 
We estimated this noise ceiling by constructing the sampling distribution of the Spearman-Brown-corrected correlation between error patterns computed on randomly partitioned halves of the human data.

When restricting estimates of human-model consistency to the items for which we obtained valid responses, we found that all models across all six tests consistently fell far short of the human noise ceiling, with different models being closer to that ceiling for different tests. 
\gptv{} was the closest to humans based on performance for GGR (\ci{0.21}{0.020}{0.45}; noise ceiling: \ci{1.00}{1.00}{1.00}), HOLF-Multi (\ci{0.63}{0.50}{0.74}; noise ceiling: \ci{0.99}{0.98}{0.99}), and ChartQA-Human (\ci{0.43}{0.29}{0.56}; noise ceiling: \ci{0.89}{0.80}{0.94}). 
\llavaThirteen{} was closest for VLAT (\ci{0.32}{0.17}{0.47}; noise ceiling: \ci{1.00}{1.00}{1.00}). 
\pix{} was closest for CALVI (\ci{0.51}{0.25}{0.78}; noise ceiling: \ci{0.99}{0.99}{1.00}). 
Finally, \llavaThirtyFour{} was closest for HOLF (\ci{0.42}{0.34}{0.50}; noise ceiling: \ci{0.89}{0.88}{0.91}). 

When estimating human–model consistency across all items, including those where a model produced only invalid responses, \gptv{} was the closest to \humans{} on each assessment. 
Nevertheless, it fell short of the human noise ceiling in every test, including GGR (\ci{0.21}{0.04}{0.42}), VLAT (\ci{0.38}{0.19}{0.57}), CALVI (\ci{0.31}{0.12}{0.50}), HOLF (\ci{0.44}{0.37}{0.50}), HOLF-Multi (\ci{0.55}{0.43}{0.65}), and ChartQA-Human (\ci{0.34}{0.22}{0.45}).


Taken together, these findings suggest that when scrutinizing the patterns of performance from models and humans more comprehensively, \revtwo{the vision-language models we evaluated generated error patterns that are reliably distinguished from those produced by humans.} 
While a proprietary system like \gptv{} was most aligned with human behavior among the models in our suite, open-source models such as \llavaThirtyFour{} and \pix{} did not necessarily lag far behind. 
These findings suggest promising opportunities for developing human-aligned models of visualization understanding. 

\section*{Discussion}
A key challenge at the intersection of cognitive science and education is to elucidate the perceptual and cognitive processes that enable understanding of symbolic displays of information, including data visualizations.
To help resolve this question, here we asked to what degree vision-language models, an emerging class of AI systems that can operate over both text and images \cite{lu2023mathvista,yue2024mmmu,huang2024pixels,li2024visualization,pandey2025benchmarking,dong2025probing,bendeck2024empirical,tartaglini2025diagnosing}, might provide the basis for developing computational cognitive models of human visualization understanding. 
We constructed a suite of visualization literacy benchmarks, \texttt{CHART-6}, which combines six tests that were developed independently by researchers across different disciplinary communities to holistically measure data visualization understanding.
This suite included five assessments intended to measure data visualization understanding in humans, GGR, VLAT, CALVI, HOLF, \& HOLF-Multi, as well as a representative subset of items from ChartQA, a commonly used benchmark that was developed to measure these skills in AI systems. 
We evaluated a set of eight \revtwo{performant} vision-language models and compared these models' performance to that of human participants.
Even when considering only valid responses from models (and thereby, if anything, \textit{overestimating} their performance), we found that models still performed worse than human participants, on average. 
Even so, the degree to which humans struggled or succeeded on a given type of question moderately predicted \gptv{}’s (the most performant model) performance on that same question type, suggesting that the current best AI systems show convergence with human patterns of behavior at a broad level.
Nevertheless, no model produced patterns of responses that approached the human noise ceiling (representing the degree of reliability in the current study's estimates of average human response patterns), indicating that there remains a meaningful gap to close before we can be confident that any vision-language model is capable of performing these tasks in a ``human-like'' manner.

Our results contribute to a growing body of cognitive-AI benchmarking efforts that employ large-scale controlled experimentation to reveal gaps between humans and AI systems on a common set of real-world tasks involving complex, naturalistic inputs \cite{mukherjee2024seva, bear2021physion, shu2021agent, fel2022harmonizing,marjieh2024large, mukherjee2024large}.
Our work is also distinct from similar efforts aimed at evaluating data visualization literacy in VLMs \cite{pandey2025benchmarking, bendeck2024empirical, mukherjee2025encqa, dong2025probing, wang2024charxiv, masry2022chartqa, wu2024chartinsights} in several key ways. Prior work has either (1) evaluated models on a small set of items or assessments, (2) primarily presented results focused on proprietary closed models, (3) did not report human performance on test items, or (4) focused solely on accuracy without looking at response \textit{patterns} across items, or some combination of the above.
We build on this prior work and address these limitations by evaluating a mix of open and closed models on a large set of items spanning different tasks and visualization types and developing a protocol that both evaluates overall model performance and similarity in response profiles between models and humans.
Taken together, our findings suggest that while vision-language models show promise as a class of models that can reason over a broad class of visualization and question types, there remain opportunities to improve their alignment with human behavior, which would enhance their value as potential scientific models of the computations involved in visualization understanding.

An outstanding question concerns where the identified gaps between models and humans come from and how to close them.
Data visualization literacy is often acquired by humans through formal education and training.
While modern vision-language models are trained on very large datasets that likely include data visualizations \cite{laurenccon2024matters}, they generally do not engage with these inputs or receive social feedback in the ways that human learners do \cite{gweon_rsta_2023, alper2017visualization, peppler2021cultivating}.
\revtwo{Indeed, while more recent instantiations of VLMs, including ``frontier'' systems such as Claude Fable \cite{anthropic2026fable} and GPT-5.6 \cite{openai2026gpt56}, are able to correctly solve many complex data visualization tasks, without conducting careful comparisons to human behavior it remains unclear to what degree these systems solve data visualization problems in human-like ways.}
An important future direction will be to uncover the aspects of human learning environments that are critical for observing robust acquisition of these skills in humans, and explore to what degree these insights can be leveraged to develop more robust and sample-efficient artificial learning systems beyond current pre-training strategies \cite{gupta2024enhancing}. 
Pursuing this direction might not only help close the quantitative alignment gap but potentially mitigate qualitative differences in behavior between vision-language models and humans. 
For example, text-based annotations embedded in data visualizations seem to influence model performance \cite{wu2024chartinsights, xu_chartbench_2024} to a greater degree than they do human performance \cite{stokes2023role}, although more direct comparisons between models and humans, similar to that in the present work, is needed.
Moreover, other work suggests that vision-language models often fail to detect visual properties that are generally salient to humans, such as intersections between lines, overlap between shapes, and the number of simple visual elements in a scene \cite{rahmanzadehgervi2024vision} --- all foundational abilities needed to succeed on the visualization understanding tasks investigated in the present work. 
One possibility is that the gaps between human and model performance on the data visualization understanding tasks in \texttt{CHART-6} can be explained, at least in part, by general limitations in models' visual processing capabilities.
Future work should seek to test this claim by concurrently evaluating models on perceptual tasks \cite{fu2024blink, hendrycks2020measuring, mukherjee2025encqa} and data visualization-understanding tasks.
 In addition, it is possible that the gaps we observed between human and model performance are at least partially attributable to differences in how information is being integrated across modalities, rather than failures in visual processing alone. If so, then interventions that enable the model to integrate visual and textual inputs more successfully should lead to improved model performance. Towards this end, it would also be fruitful to investigate the causal impact of augmenting base vision-language models with better reasoning mechanisms \cite{wei2022chain, zelikman2022star}.

Another important future direction will be to develop more unified measures of data visualization literacy.
Currently, the landscape of assessments and benchmarks for measuring these skills is fragmented, and there is a lack of consensus regarding the key components of data visualization literacy and exactly how to measure them \cite{verma2025measuring, borner2016investigating,boy2014principled,lee2016vlat,ge2023calvi,huey2023communicative,masry2022chartqa,wang2024charxiv, zeng2024advancingmultimodallargelanguage}. 
Furthermore, there might also be important aspects of data visualization literacy that are not well captured by existing measures\cite{brockbank2025evaluating}. 
Many benchmarks used in the machine learning literature \cite{lu2023mathvista, hendrycks2020measuring, wang2024charxiv, wu2024chartinsights, cui2024promises} contain a large number of graphs that are similar to those that can be found in real-world settings, yet the questions accompanying them are often simpler than would be expected for a comprehensive measure of data visualization understanding.
Meanwhile, assessments of data visualization literacy designed for humans often contain fewer items, but tap into a broader array of skills \cite{galesic2011graph, lee2016vlat, ge2023calvi}. 
Future work could analyze the properties of existing measures \cite{brockbank2025evaluating} and leverage the resulting insights to develop scalable procedures for developing more comprehensive measures \cite{cui2024promises}.
Adaptive testing methods might also be used to more efficiently administer these comprehensive tests to humans \cite{cui2023adaptive}.


While our study included models with different architectures, training objectives, and datasets, these factors were not under experimental control, and our suite represented only a modest subset of vision–language models capable of data visualization understanding. 
Thus, we cannot draw strong inferences about the causes of the behavioral differences we observed, or about their generalizability to vision–language models \textit{writ large}.
In addition, we limited our study to assessments for which there was human behavioral data readily available \cite{galesic2011graph, lee2016vlat, ge2023calvi, huey2023communicative, huey2024adaptive}. 
However, there are other assessments that would otherwise be valuable to include in a more comprehensive human-centered evaluation of vision-language models \cite{cui2024promises, boy2014principled, wainer1980test, cui2023adaptive}. 
Despite these limitations, a core contribution of our work is to position data visualization literacy --- measured with validated human assessments --- as a key target for vision–language models, and to establish a protocol for `species-fair' \cite{firestone2020performance} comparisons between humans and models that can be easily deployed with the release of newer iterations of frontier models.



Visualizations are a powerful tool for supporting discovery and communication of patterns in quantitative data.
We envision \texttt{CHART-6} being used to track the progress of artificial systems towards achieving human-like behavior on tasks involving data visualizations, and thus a procedure for identifying promising systems for further investigation as candidate computational models of the cognitive operations involved in these tasks.
Here we found that many current vision-language models show promise, but still fall short of providing a strong foundation for developing cognitive models.
As progress is made on this front, we believe it to be likely that AI systems displaying more human-like understanding of visual, linguistic, and mathematical concepts could be used to design more effective STEM learning environments and tools to support scientific communication.


\section*{Methods}
\begin{figure*}[ht!]
    \centering
    \includegraphics[width=0.99\linewidth]{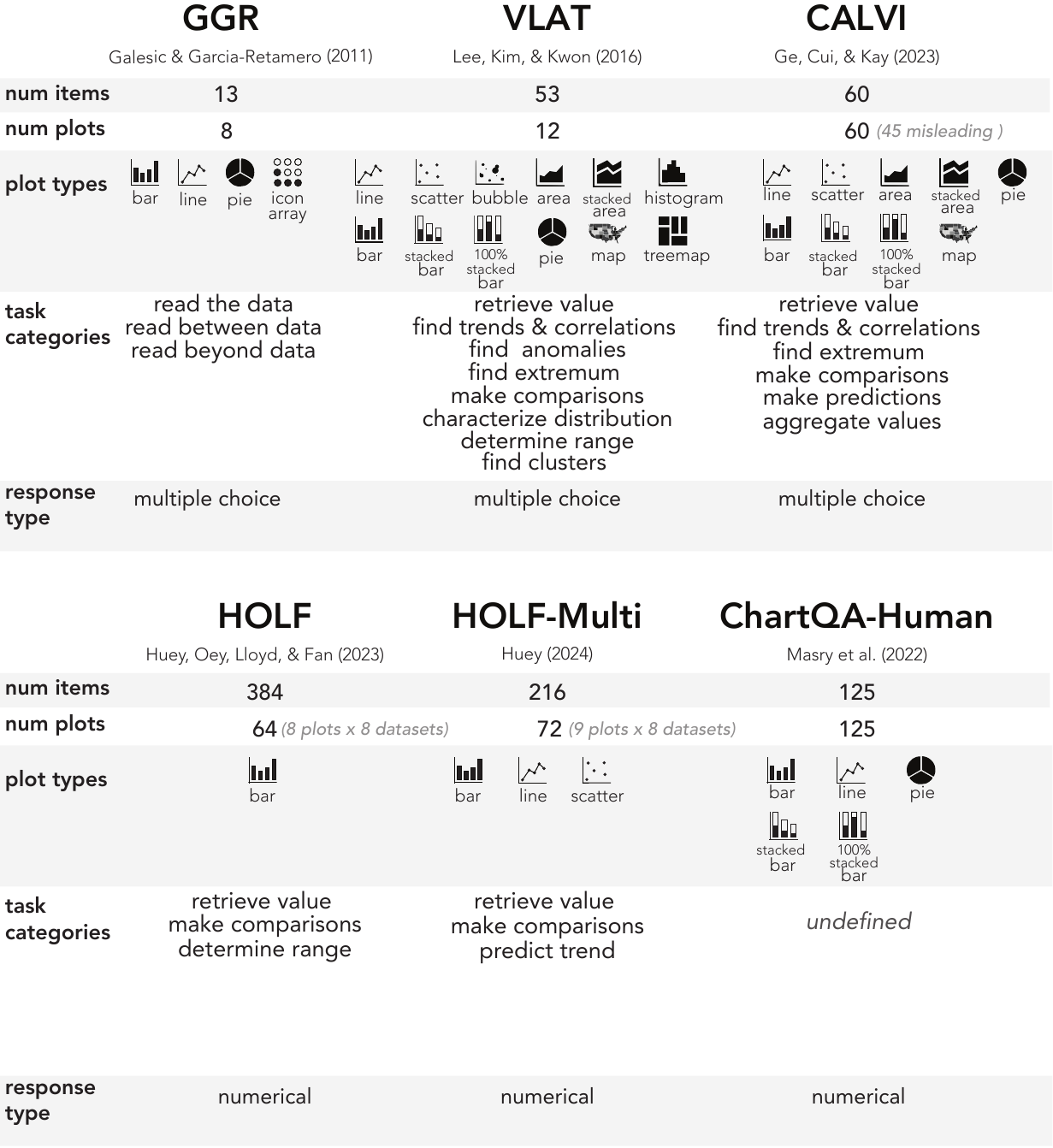}
    \caption{We present \texttt{CHART-6} (\textit{Comparative Human-AI Graphical Reasoning Tests}), a human-centered suite of data visualization understanding benchmarks, to assess how close state-of-the-art vision-language models are to achieving both \textit{human-level} performance and \textit{human-like} behavior on reasoning tasks involving data visualizations. This test suite spans a wide array of different approaches to designing such assessments, ensuring broad coverage of the skills that are considered to be important when assessing human data visualization understanding.}
    \label{fig:methods}
\end{figure*}

Progress towards computational models that emulate human understanding of data visualizations requires meeting two key methodological challenges: (1) establishing common standards by which to assess understanding of data visualizations in humans and AI systems, and (2) conducting controlled evaluations of human and AI understanding of data visualizations that support direct comparison between these two systems.
This effort follows in the tradition of recent human and AI benchmarking work in the cognitive sciences \cite{bear2021physion, mukherjee2024seva, bonnen2024evaluating, fel2022harmonizing, geirhos2018generalisation, marjieh2024large, mukherjee2024large}.

\subsection*{Test suite}
\label{sec:test-suite}

Leveraging prior work on developing tests of data visualization literacy in the psychology and visualization literature \cite{galesic2011graph, lee2016vlat, huey2023communicative, huey2024adaptive, ge2023calvi} we developed a diverse test suite that provides broad coverage of the skills that are considered to be important when assessing data visualization literacy in humans (Figure~\ref{fig:methods}).

All of these tests consist of a series of test items, each presenting an image of a data visualization paired with a question posed in natural language.
Three tests consist primarily of multiple-choice questions, requiring a response that matches one of several provided options. 
The remaining three tests consist of questions requiring a numerical response.
Since many tests had multiple questions paired with a given visualization, we refer to each unique visualization-question pair as a test `item' in each of the tests. Below, we provided a brief description of each of the six tests included in \texttt{CHART-6}.

\paragraph{GGR.} GGR is a 13-item test containing three bar plots, three line plots, an icon array, and a pie chart \cite{galesic2011graph}. 
The test was designed to probe a compact hierarchy of abstract abilities, progressing from ``reading the data'' to ``reading between the data'' to ``reading beyond the data'' \cite{friel2001making}. 
Nine of the test items required a numerical response and four of them were multiple choice. 
While the answers for several items are numeric, since the designers of the test assessed performance by computing the proportion of responses that were exact matches to the true answer, we also treated this test as one whose answers were `multiple-choice'.

\paragraph{VLAT.} The Visualization Literacy Assessment Test (VLAT) is a 53-item test containing 12 graph types \cite{lee2016vlat} --- line chart, bar chart, stacked bar chart, normalized stacked bar chart, pie chart, histogram, scatter plot, bubble chart, area chart, stacked area chart, choropleth map, and tree map --- each generated using data obtained from news articles.
VLAT groups items into more concrete tasks than in GGR, including questions that involve: retrieving values, finding extrema, finding anomalies, making comparisons, determining ranges, finding correlations \& trends, and finding clusters. 
All of the test items are multiple choice (34 items with four options; 3 with three options; 16 were True/False).

\paragraph{CALVI.} The Critical Thinking Assessment for Literacy in Visualizations (CALVI) is a 60-item test focusing on the ability to compensate for misleadingly constructed data visualizations, such as the use of inappropriate scale ranges or unconventional scale directions \cite{ge2023calvi}. It is composed of 45 items which feature such misleading visualizations, enabling direct comparison between human and model behavior in cases where many humans are expected to fail.
All of the test items in CALVI require multiple-choice responses.

\paragraph{HOLF.} HOLF is a 384-item test containing 64 bar plots procedurally generated from eight real-world datasets. 
Each chart was paired with six different questions measuring the ability to retrieve values, make comparisons, and determine ranges, yielding 48 unique questions in total. 
While in VLAT and GGR each plot is paired with an uneven number and variety of types of questions, in HOLF each plot was paired with all six question types, making it possible to disentangle the impact of various plot attributes from properties of the underlying dataset. 
This test was originally used in controlled laboratory settings to characterize human judgments concerning which of several plots would be most informative to other people for answering a particular question \cite{huey2023communicative}.
\paragraph{HOLF-Multi.} HOLF-Multi is a 216-item extension of HOLF containing 72 bar, line, and scatter plots \cite{huey2024adaptive}.
What distinguishes HOLF-Multi from HOLF is a larger variety of graph types.
These plots were generated from the same eight datasets as in HOLF, and each plot was paired with 3 questions, yielding a total of 24 unique questions. 

\paragraph{ChartQA-Human.} ChartQA \cite{masry2022chartqa} is a data visualization understanding benchmark containing plots obtained from various web sources such as Statista and Pew Research. 
An initial set of questions about them was generated by a combination of human participants and language models, which was then refined by the benchmark developers. 
Vision-language models are routinely evaluated on the test split of this benchmark, which \revtwo{requires providing natural language responses to} 2,490 questions pertaining to 1,509 plots.
\revtwo{In CHART-6, we subsampled items from the full dataset with two key considerations in mind. 
First, we only included items that clearly required numerical responses to make the space of responses more tractable for analysis.
Second, since we needed to collect human responses for these items, we subsampled to a set of items small enough that each crowdworker could complete a reasonable number of trials without requiring an impractically large number of participants.
}
We constructed ChartQA-Human by sampling a random subset of 125 items from the ChartQA test set such that different types of graphs, data sources, and question styles (i.e., human-written vs. template-based) appeared in roughly equal proportion to their relative frequency in the full set of ChartQA items. 
\revtwo{ChartQA-Human, thus, constitutes a subset of the full benchmark that preserves the diversity of visualization and question types.}
\subsection*{Task categories}
\begin{table*}[ht!]
\centering
\renewcommand{\arraystretch}{1.2}
\begin{tabularx}{\textwidth}{
>{\raggedright\arraybackslash}l 
>{\raggedright\arraybackslash}X 
>{\raggedright\arraybackslash}X 
>{\raggedright\arraybackslash}X}
\toprule
\textbf{Test} & 
\textbf{Value Identification} & 
\textbf{Arithmetic Computation} & 
\textbf{Statistical Inference} \\
\midrule
\textbf{GGR} & read the data & read between the data & read beyond the data \\ 
\midrule
\textbf{VLAT} & retrieve value, find extremum, determine range & make comparisons & find correlations/trends, characterize distribution, find anomalies, find clusters \\
\midrule
\textbf{CALVI} & retrieve value{
\color{Navy}, find extremum} & make comparisons & find correlations/trends, make predictions, aggregate values \\
\midrule
\textbf{HOLF} & retrieve value & make comparisons, determine range & --- \\
\midrule
\textbf{HOLF-Multi} & retrieve value & make comparisons & predict trend \\
\midrule
\textbf{ChartQA-Human} & --- & --- & --- \\
\bottomrule
\end{tabularx}
\caption{The relationship between the original task categories and the common set of task categories that can be applied across all six tests. Each row contains the names of the task categories originally defined in each test.}
\label{table:task_mapping}
\end{table*}

Because these six tests were developed independently of one another, they used ways of organizing items into task categories that were not commensurate with one another (e.g., ``find trends \& correlations'' and ``read beyond the data'').
To conduct analyses that spanned these different tests, we defined a common set of task categories that could be applied to all tests (Table~\ref{table:task_mapping}): \emph{value identification}, where participants retrieve an individual value appearing in a plot (e.g., finding the maximum value); \emph{arithmetic computation}, where participants are expected to perform simple arithmetic operations over multiple values displayed in the plot (e.g., finding the average of two values); and \emph{statistical inference}, where participants must estimate latent parameters in a statistical model based on the values shown (e.g., judge the strength of trends or presence of clusters).
The only exception was ChartQA-Human which did not initially specify any task categories to organize the test items it contains.

\subsection*{Measuring data visualization understanding in humans}

\begin{figure*}[ht!]
    \centering
    \includegraphics[width=0.95\linewidth]{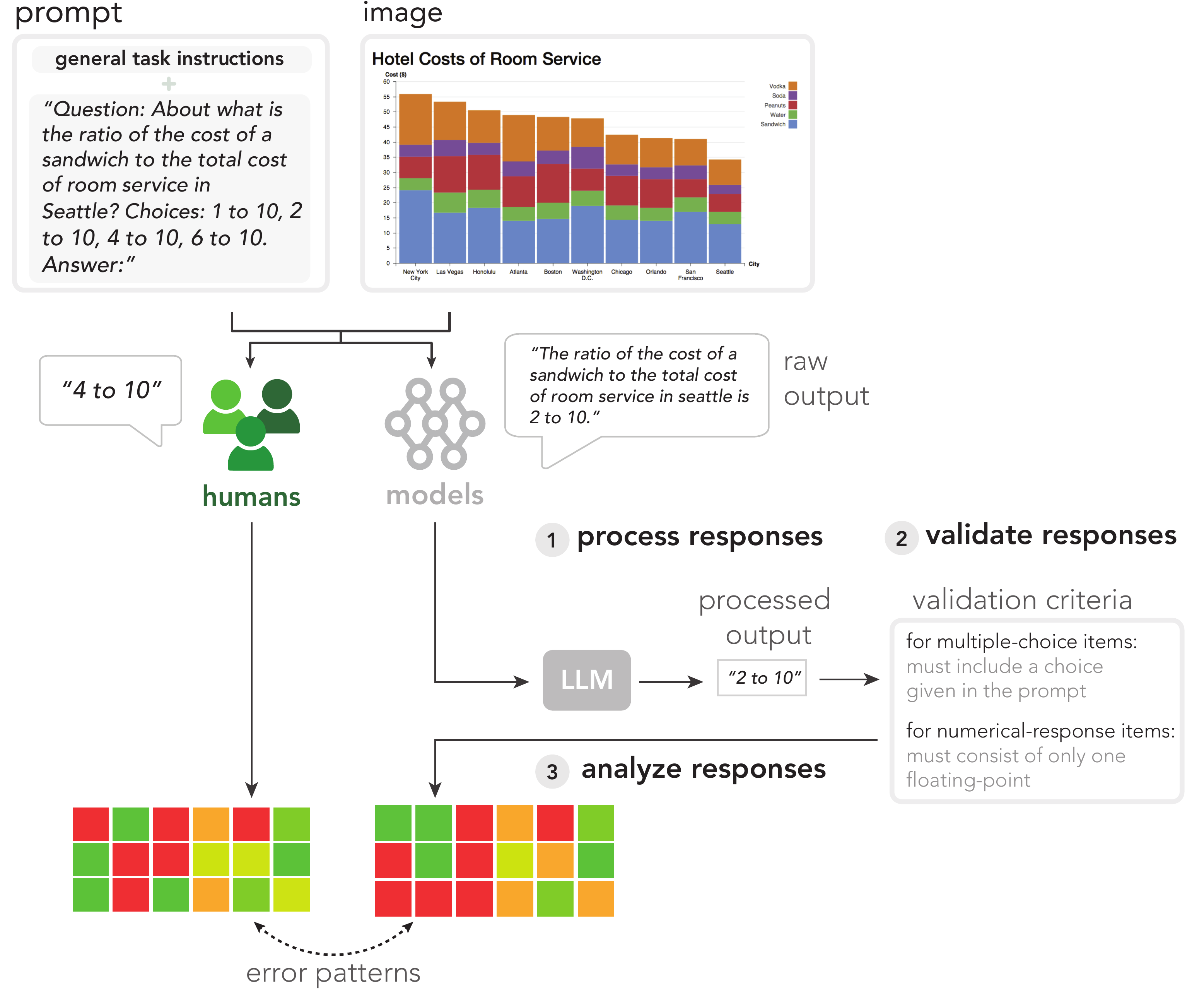}
    \caption{Procedure for processing and validating model responses for comparison to human responses. All vision-language models were presented with every test item 10 times. Each test item consisted of an image containing a data visualization and a question accompanying it, preceded by general task instructions. The raw output generated by each model was then processed independently by a different large-language model to extract the response in the correct format. These processed outputs were then scored and the pattern of errors compared to human error patterns.}
    \label{fig:response-pipeline}
\end{figure*}

Where available, we leveraged existing human behavioral data, and where necessary, collected new data by conducting studies with human participants.

\paragraph{GGR and VLAT.} Data were collected in a previous study with 1,135 human participants recruited via Prolific \cite{lloyd2023graph}.
Each participant was asked to complete both of these tests in a single session  
with test order randomized across participants.

\paragraph{CALVI.} Data were collected in a previous study with 497 participants
\cite{ge2023calvi}. 
Participants were recruited via Prolific and given a 30-item test: 15 were randomly sampled from the set of 45 misleading items, while the other 15 were always the same set of non-misleading items. Note that, unlike a human participant, each VLM was evaluated on all 60 items instead of a random sample. This means that, while it may not be appropriate to compare the performance of individual humans to a VLM, in aggregate all VLMs and humans were evaluated on the same questions.

\paragraph{HOLF and HOLF-Multi.} Data were collected in a previous study with 531 participants on HOLF 
and 1,743 participants on HOLF-Multi \cite{huey2023communicative, huey2024adaptive}. 
In both studies, each participant was presented with eight items drawn from the full set of test items, such that they only saw one plot and question pertaining to each of the eight datasets. 

\paragraph{ChartQA-Human.} 
We recruited 50 participants via Prolific in the present study to complete ChartQA-Human, a 125-item representative subset of the ChartQA benchmark.
Each participant completed a set of 25 items sampled at random from the full set of 125 items. \revtwo{Every item was answered 10 times.}
Participants provided informed consent. All methods were carried out in accordance with relevant guidelines and regulations. All experimental protocols were approved by a named institutional and/or licensing committee, the Stanford Institutional Review Board, and participants were compensated for their time (\$15.50 per hour).

\subsection*{Measuring data visualization understanding in models}
\label{sec:models}

\paragraph{Model suite}
To determine which vision-language models to include in our evaluation, we prioritized those that achieved strong performance on other benchmarks that involve reasoning over visual and linguistic inputs \cite{li2023seed, yue2024mmmu}.
In addition, to improve the robustness of our findings, we sought to include a suite of models that was reasonably diverse and representative of current modeling approaches with respect to architecture, size, and pre-training protocol.
We selected eight models in total, which included three pairs of models that shared similar architectures and training regimes.
\flanXL{} and \flanXXL{} used the 4B-parameter (FlanT5-XL) and 11B-parameter (FlanT5-XXL) versions of the FlanT5 language model respectively \cite{chung2024scaling}. Both models used the same BLIP-2 pre-training regimen \cite{li2023blip}.
Similarly, \llavaSeven{} and \llavaThirteen{} used the same  CLIP-ViT-L-336px vision encoder and the 7B-parameter and 13B-parameter version of the Vicuna language model respectively. Both models were trained using the LLaVA-1.5 framework \cite{liu2024improved, radford2021learning}.
\matcha{} \cite{liu2023matcha} augments the pre-training of \pix{} \cite{lee2023pix2struct} with additional tasks intended to enhance its general visual and quantitative reasoning performance.
We also included \llavaThirtyFour{} \cite{liu2024improved}, which uses the 34B-parameter version of the Yi language model \cite{ai2024yiopenfoundationmodels} trained using the LLaVA-1.5 framework.
Finally, we evaluated \gptv{}, a highly performant proprietary model 
\cite{openai2023gpt4}.

\paragraph{Extracting model output} 
Each model was evaluated on all 851 test items from GGR, VLAT, CALVI, HOLF, HOLF-Multi, and ChartQA-Human. 
The input to models consisted of two components: (1) an image containing a data visualization and (2) a text prompt containing a question about the visualization written in English. 
General instructions describing the nature of the task were prepended to each question.
In addition, all prompts were formatted to match the model-specific prompts used during training (e.g., prepending the word \textit{Question:} before each question; see Appendix for details).

To assess the test-retest reliability of responses generated by models, we presented each test item 10 times to every model, yielding a total of 8,510 responses per model. We selected 10 runs to align with the upper end of the range used in comparable prior work\cite{pandey2025benchmarking,bendeck2024empirical}. Furthermore, keeping the response for each run instead of selecting the best run allows a weighted estimate of a model's accuracy on a particular item.

We explored commonly used strategies for improving the diversity and fluency of model outputs, including nucleus sampling \cite{holtzman2019curious, renze2024effect}, a decoding procedure wherein sampling is performed over the smallest possible set of words whose cumulative probability exceeds a probability threshold, \textit{top-p}. 
Specifically, we identified the combination of \textit{top-p} and \textit{temperature} values for each model that produced the best performance on one test (in our experiments, VLAT), and then used these same model-specific \textit{top-p} and \textit{temperature} values for the remaining tests. 

\paragraph{Processing model output.} Determining whether a model had correctly answered a question usually required further processing and validation (Figure~\ref{fig:response-pipeline}; see Appendix for details). 
For instance, several models produced verbose responses that did not conform to any of the required response formats (i.e., multiple choice, True/False, numerical response). 
In particular, \llavaSeven{} and \llavaThirteen{} often returned the full input prompt as part of its response, so we applied further processing to excise the prompt from any responses that included them. 
Following prior work \cite{yue2024mmmu}, we also used GPT-4 to extract only the relevant information in the correct response format from the raw model output. 
For items that required a floating-point answer, any strings prefixed or suffixed to the floating-point value (e.g. ``\$'' in ``\$3.27'' or ``cm'' in ``4cm'') were removed. All prompt templates were adopted from the MathVista benchmark\cite{lu2023mathvista}.


\paragraph{Setting model hyperparameters.} 
\begin{table}[ht!]
\centering
\begin{tabular}{lcc}
\toprule
\textbf{Model} & \textbf{\textit{top-p}} & \textbf{\textit{temperature}} \\ 
\midrule
\flanXL{} & 0.6 & 1.0 \\ 
\flanXXL{} & 1.0 & 1.0 \\
\llavaSeven{} & 0.4 & 1.0 \\
\llavaThirteen{} & 1.0 & 0.4 \\
\llavaThirtyFour{} & 1.0 & 0.4 \\
\pix{} & 0.8 & 1.0 \\
\matcha{} & 0.4 & 1.0 \\
\gptv{} & 1.0 & 0.2 \\
\bottomrule
\end{tabular}
\caption{\textit{Top-p} and \textit{temperature} hyperparameters used in the current model evaluation study.}
\label{table:hyperparams}
\end{table}

The \textit{max\_new\_tokens} parameter for all models was set to 270, a relatively high value in order to reduce the likelihood of obtaining a prematurely truncated response. 
We conducted a search over possible combinations of \textit{temperature} and \textit{top-p} parameters that maximized each model's performance on VLAT, then used these values when evaluating that model on the remaining assessments.
We considered \textit{temperature} values of 0.2, 0.4, 0.6, 0.8, and 1.0 when \textit{top-p} was set to 1.0; and \textit{top-p} values ranging between 0.2, 0.4, 0.6, 0.8, and 1.0 where \textit{temperature} was set to 1.0 (Table~\ref{table:hyperparams}).

\subsection*{Statistical analyses}

Overall, our statistical analyses aimed to account for reliable variation in model behavior across vision-language models and human participants. 
We additionally explored the contribution of other factors, including test type, question type, and graph type.
Towards this end, we fit generalized mixed-effects linear regression models to assess the relative contribution of each of these factors in predicting model and human responses.
We used non-parametric resampling methods to provide quantitative estimates of effect sizes for each factor. 

\paragraph{Linear models}
We constructed linear regression models to assess the effect of different predictors (i.e., graph type, task type, model) on visualization understanding performance. 
We used nested model comparisons as our general approach to hypothesis testing as it provides a unified framework that goes beyond the narrower set of cases considered by traditional hypothesis tests (e.g., \textit{t}-tests, ANCOVAs).

For example, to assess whether different vision-language models reliably varied in performance, we fit a mixed-effects logistic regression model predicting whether a response was correct or incorrect from model type, fitting random intercepts for each test item. 
We then compared the fit of this model to a null model that included only the random-effects term for item.
In more targeted analyses comparing \gptv{} and \humans{} across items involving different types of graphs, we fit a mixed-effects linear regression model predicting proportion correct from ``agent type'' (i.e., all models and \humans{}), graph type, and their interaction, with variation across individual items modeled using random intercepts. 
To assess the degree to which any performance gap between \gptv{} and \humans{} differed across items involving different graph types, we compared the above model to one without the interaction term. 
We conducted an analysis following the same structure to compare performance by \gptv{} and \humans{} across different task categories.  



\paragraph{Confidence intervals}

To estimate uncertainty in our point estimates of performance, we constructed 95\% confidence intervals using bootstrap resampling. 
For each model, we resampled items with replacement 1,000 times, each time re-computing performance and retaining only items for which valid responses were ever generated. 
To estimate differences between any two groups, we constructed 95\% confidence intervals based on the sampling distribution for the difference between the point estimates for each group derived on each bootstrap resampling iteration.

\section*{Acknowledgements}
We thank members of the Cognitive Tools Lab for their input and feedback. 

\section*{Author contributions statement}

All authors contributed to the writing, reviewing, and editing of the manuscript. A.V., J.E.F., and K.M. contributed to software development and statistical analysis. J.E.F., E.K., and C.P. contributed to supervision and funding acquisition.

\section*{Data availability statement}
All experimental materials, data, and analysis code are available at: \\ \url{https://osf.io/e25mu/?view_only=399daff5a14d4b16b09473cf19043f18}.
Task stimuli and assessment items are mirrored at  \url{https://github.com/cogtoolslab/human-vlm-datavis-eval}.

\section*{Funding}
This work was supported by NSF CAREER Award \#2436199, NSF DRL \#2400471, and awards from ONR Science of Autonomy, the Stanford Human-Centered AI Institute (HAI), and Stanford Accelerator for Learning.

\bibliography{sections/bibliography}





\end{document}